\documentclass[aps,prd,onecolumn,amsmath,nofootinbib,preprintnumbers]{revtex4}

\usepackage{amsmath}
\usepackage{graphicx}
\usepackage{dcolumn}
\usepackage{bm}
\usepackage{amssymb}
\usepackage{latexsym}

\begin{document}

\preprint{}

\title{Scalar Perturbation and Stability of Ricci Dark Energy}

\author{Chao-Jun Feng}
\email{fengcj@itp.ac.cn} \affiliation{Shanghai United Center for Astrophysics(SUCA), Shanghai Normal University, 100
Guilin Road, Shanghai 200234,China} \affiliation{Institute of Theoretical Physics, CAS, Beijing 100190, P.R.China}
\author{Xin-Zhou Li}
\email{kychz@shnu.edu.cn} \affiliation{Shanghai United Center for Astrophysics(SUCA), Shanghai Normal University, 100
Guilin Road, Shanghai 200234,China}


\begin{abstract}
The Ricci dark energy (RDE) proposed to explain the accelerating expansion of the universe requires its parameter
$\alpha < 1$, whose value will determine the behavior of RDE. In this Letter, we study the scalar perturbation of RDE
with and without matter in the universe, and we find that in both cases, the perturbation is stable if $\alpha> 1/3$,
which gives a lower bound for $\alpha$ theoretically.
\end{abstract}


\maketitle

Observations like Type Ia supernovae, CMB and SDSS et al. have strongly confirmed that our universe is accelerated
expanding recently. However, since ordinary matter such as stars always attract each other due to the Newton's gravity,
the universe can only be decelerated expanding. Thus, there must be an unknown energy component living in the universe,
and people often call it dark energy. Experiments have indicated there are mainly about $73\%$ dark energy and $27\%$
matter components in the recent universe, but so far people still do not understand what is dark energy from
fundamental theory. The best candidate seems the cosmology constant including the vacuum energy, but it suffers the
fine-tuning and coincidence problems. In order to alleviate these problems, a lot of dynamic dark energy models have
been built, such as quintessence, phantom, quintom models which are basically scalar field models. Another way to
explain the accelerating is to modify the Einstein gravity theory like the famous $f(R)$ theory and DGP model etc..

Actually, the cosmology constant (or dark energy) problem is in essence an issue of quantum gravity
\cite{Witten:2000zk}, since the density of dark energy is inevitably related to the large vacuum energy density of the
quantum field theory without including gravity. Considering the gravity effects, there may be some regions in which the
field theory can be valid. The holographic principle regards the black hole as the object with maximum entropy in a
given region, and from statistical physics, the entropy is a extensive quantity which proportional to the volume of
such region, while the black hole' entropy is proportional to the area of its surface, so in the field theory, there
should exists a infrared (IR) cutoff, beyond which the field theory will be failed. However, such constraint seems a
little bit loose, because it includes the black hole state in the field theory. To avoid the existence of such states,
Cohen et al.\cite{CohenZX:1999} suggested that in a given region with length scale $L$, the field's energy should be
bounded by the black hole's, i.e. $\rho L^3 \le LM^2_{pl}$, where $\rho$ is the total energy density within the region
and $M_{pl} = G^{-1/2}$ is the Planck mass. Applying the holographic principle to cosmology, Li \cite{Li:2004rb} has
proposed the holographic dark energy model, in which the energy density of dark energy is $\rho = 3c^2M^2_{pl}L^{-2}$,
namely it saturates the bound. He finds that when $L = R_h$, which is the future event horizon, this model will be
consistent with observations and meanwhile solves the coincidence problem.

Although the holographic model based on the future event horizon is successful in fitting the current data, some
authors asked why the current acceleration of the universe is determined by its future. Actually, the future event
horizon is not the only choice for the holographic dark energy model. Also motivated by the holographic principle, Gao,
et al.\cite{Gao:2007ep} have proposed the Ricci dark energy (RDE) model recently, in which the future event horizon
area is replaced by the inverse of Ricci scalar, and this model is also phenomenologically viable.

Assuming the black hole is formed by gravitation collapsing of the perturbation in the universe, the maximal black hole
can be formed is determined by the casual connection scale $R_{CC}$ given by the "Jeans" scale of the perturbations.
For tensor perturbations, i.e. gravitational perturbations, $R_{CC}^{-2} = Max(\dot H + 2H^2 , -\dot H )$ for a flat
universe, where $H = \dot a/a$ is the Hubble parameter, and according to the ref.\cite{Cai:2008nk}, only in the case of
$R_{CC}^{-2} =\dot H + 2H^2$,  it could be consistent with the current cosmological observations when the vacuum
density appears as an independently conserved energy component. As we know, in flat FRW universe, the Ricci scalar is
$R=6(\dot H + 2H^2)$, which means the $R_{CC} \propto R$ and if one choices the casual connection scale $R_{CC}$ as the
IR cutoff, the Ricci dark energy model is also obtained. For recent progress on Ricci dark energy and holographic dark
energy, see ref.\cite{recentRicci}\cite{holo:recent}\cite{Zhang:constraint}. The energy density of RDE in flat universe
reads
\begin{equation}\label{energy density}
    \rho_{R} = \frac{\alpha}{2}R = 3\alpha \left(\dot H + 2H^2 \right) \,,
\end{equation}
where  we have set $8\pi G = 1$ and $\alpha$ is a dimensionless parameter which will determine the evolution behavior
of RDE.

In the following, we will investigate the scalar perturbation of RDE model. At first, we only consider the case when
the universe is dominated by RDE and analytically solve the equation of motion for the perturbation. And then we
consider the case when matter comes in, namely, the universe dominated by both RDE and matter, the equation of motion
for perturbation is too complicated to be solved analytically, but it can be solved in the long-wave and short-wave
limit. And we also solve it numerically. In both cases, it shows that, the perturbation is stable if $\alpha>1/3$  and
this result gives a lower bound for $\alpha$ theoritically.

The linear scalar perturbation of the flat FRW metric in longitudinal (Newtonian) gauge is given by :
\begin{equation}\label{linear pert}
    ds^2 =  -\left(1+2\Phi\right)d t^2 + a^2(1-2\Psi)\delta_{ij}dx^i dx^j \,,
\end{equation}
and we assume that the perturbation has spherical symmetry, namely, there is no anisotropic stress to the linear-order
of the perturbation, then from the off-diagonal $ij$ perturbed Einstein equations one can see $\Phi = \Psi$. Therefore
the perturbation of RDE is
\begin{equation}
    \delta \rho_R = \frac{\alpha}{2} \delta R = \alpha  \bigg[ \frac{\nabla^2}{a^2} \Phi - 3\ddot\Phi - 15H\dot\Phi-6(\dot H+ 2H^2)\Phi
    \bigg]\,,
\end{equation}
and the $00$ component of perturbed Einstein equations is given by
\begin{equation}\label{perturb einstein}
    2\left[-3H^2\Phi - 3H\dot\Phi+ \frac{\nabla^2}{a^2} \Phi \right] = \delta \rho_R +
    \delta \rho_m  \,,
\end{equation}
where  $\delta \rho_m$ denotes the perturbation of matter.

\begin{figure}[h]
\includegraphics[width=0.4\textwidth]{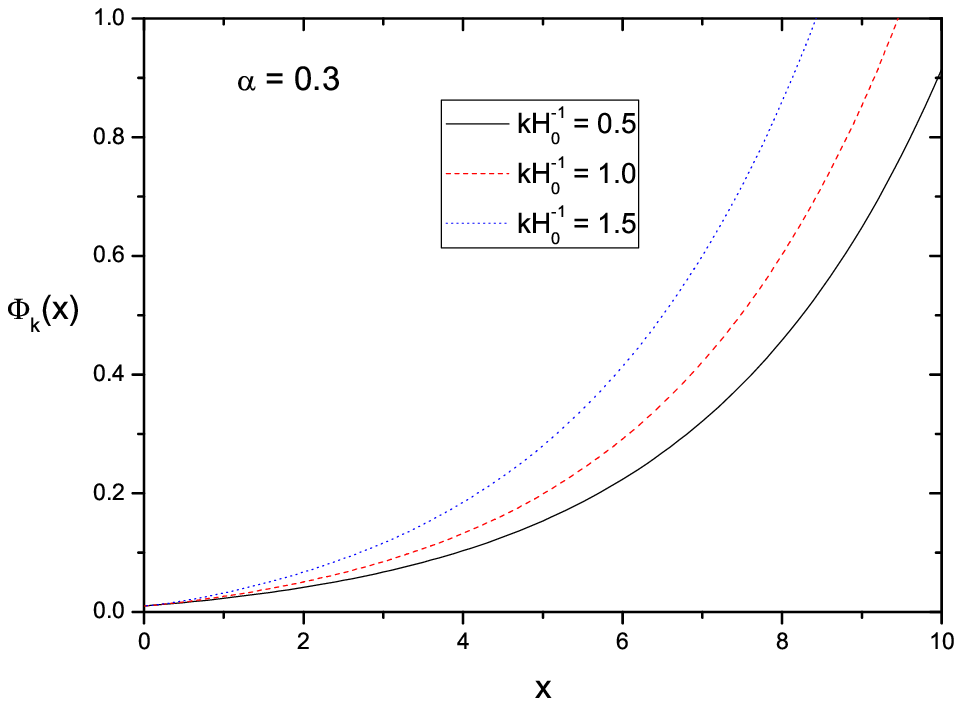}
\quad
\includegraphics[width=0.4\textwidth]{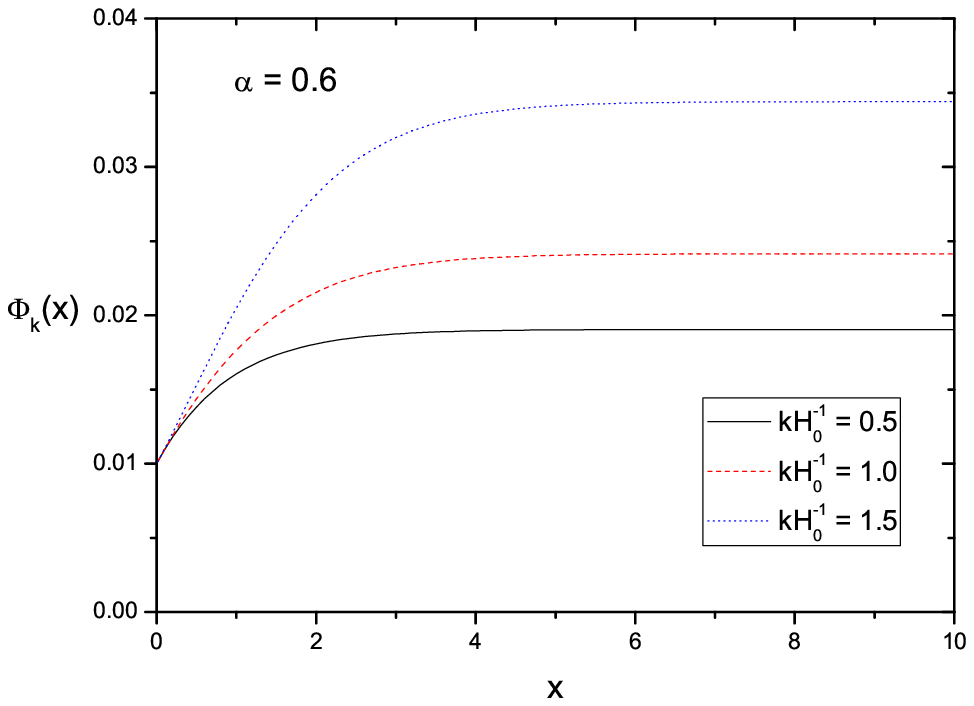}
\caption{\label{fig::PhiK}\emph{RDE without matter}: Evolution of perturbations $\Phi_k$  with $\alpha = 0.3$(left),
$0.6$(right), $kH_0^{-1} = 0.5, 1.0, 1.5$ and the initial condition $\Phi_{k0} = 0.01$, $\Phi'_{k0} = 0.01$. }
\end{figure}

\begin{figure}[h]
\includegraphics[width=0.31\textwidth]{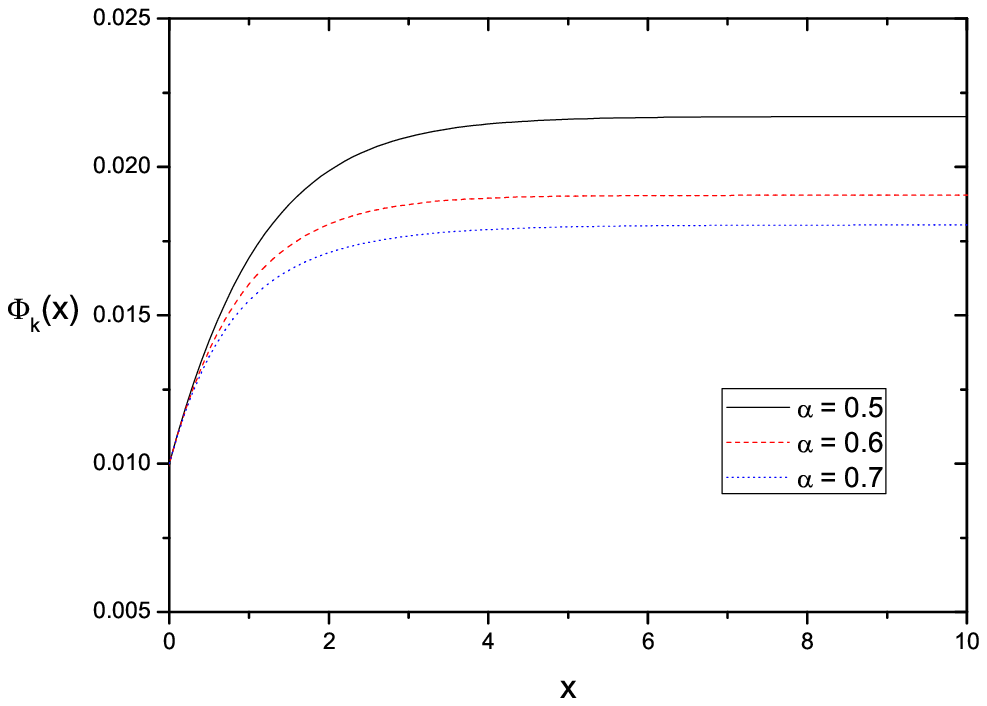}
\quad
\includegraphics[width=0.31\textwidth]{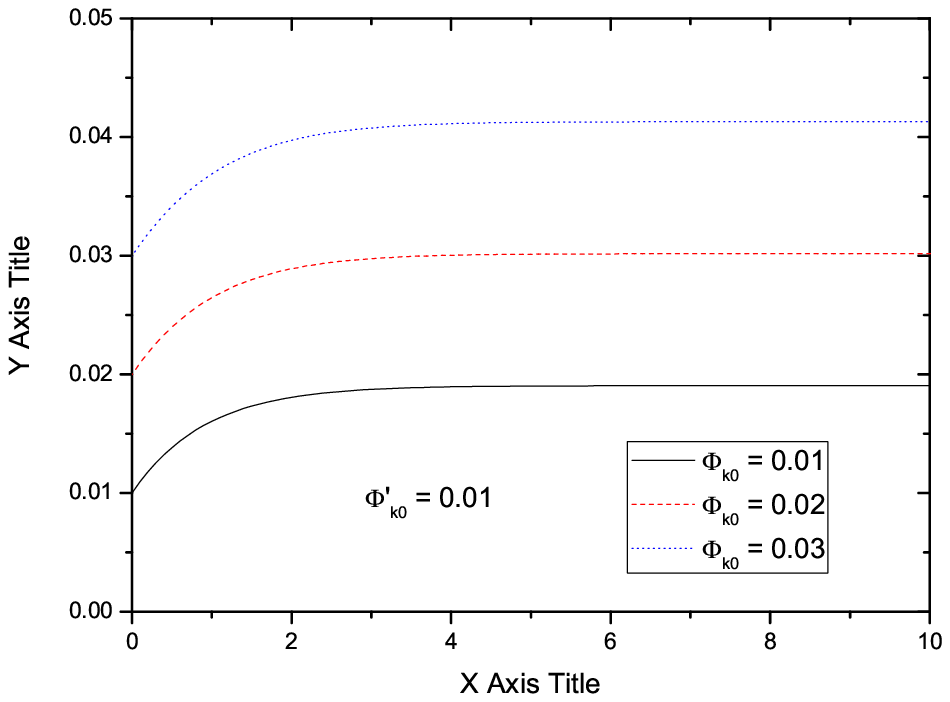}
\quad
\includegraphics[width=0.31\textwidth]{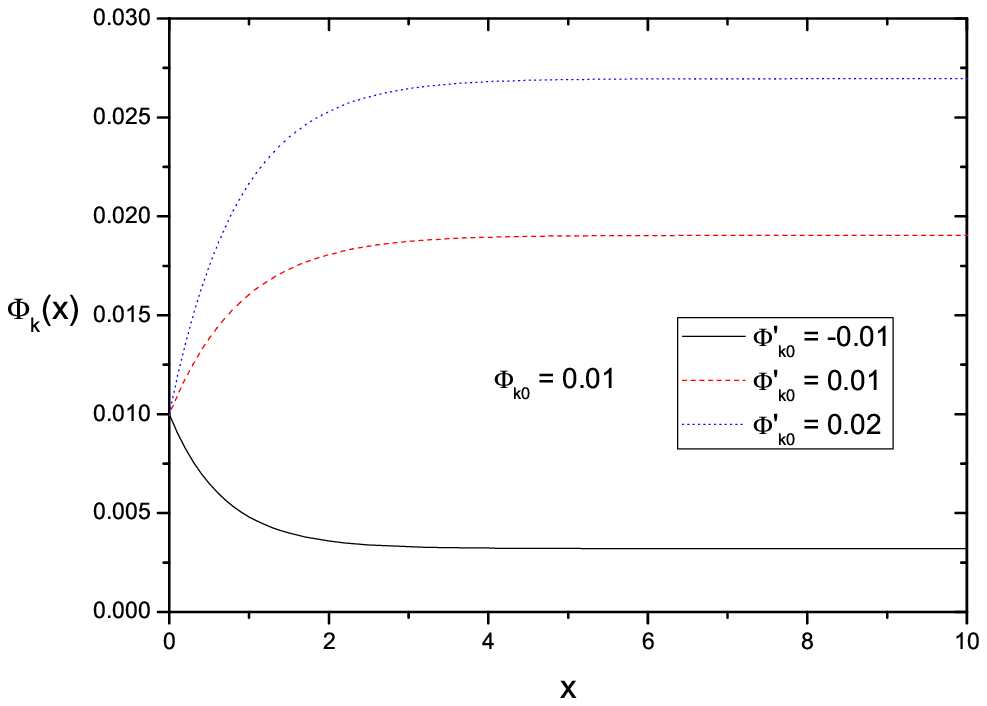}
\caption{\label{fig::PhiAlpha}\emph{RDE without matter}: The most left figure shows the evolution of perturbations
$\Phi_k$ with $\alpha = 0.5, 0.6, 0.7$ and the initial condition $\Phi_{k0} = 0.01$, $\Phi'_{k0} = 0.01$. The middle
and the most right figures show the evolution of $\Phi_k$ with $\alpha = 0.6$ and different initial conditions:
$\Phi_{k0} = 0.01, 0.02, 0.03$, $\Phi'_{k0} = 0.01$ in the middle one and $\Phi_{k0} = 0.01$, $\Phi'_{k0} = -0.01,
0.01, 0.02$ in the most right one.}
\end{figure}

\emph{RDE without matter}: First, let us consider the case when the universe is dominated by RDE only, then the
Friedmann equation reads:
\begin{equation}\label{background no matter}
    H^2 = \frac{\rho_R }{3} =  \alpha \bigg( \dot H  + 2H^2 \bigg) = \alpha \bigg[  \frac{(H^2)'}{2}  + 2H^2 \bigg]\,,
\end{equation}
where prime denotes the derivative with respect to $x \equiv \ln a$ and hereafter we set $a_0 = 1$. The solution of the
above equation is $H^2 = H_0^2e^{-2\left(2-\frac{1}{\alpha}\right)x}$ and the energy density of RDE is $\rho_R  =
3H_0^2e^{-2\left(2-\frac{1}{\alpha}\right)x}$. By using the energy conservation law we get it's equation of state as
\begin{equation}\label{eos for no matter}
    w = -1 - \frac{(\ln\rho_R)'}{3} = -\frac{1}{3}\left(\frac{2}{\alpha}-1\right) \,,
\end{equation}
and to make the universe accelerated expanding ($w<-1/3$), it requires $\alpha < 1$. Since in the linear theory of
cosmology perturbations all Fourier modes evolve independently, we will focus on each mode labeled by its comoving wave
number $k$. By using the background equation (\ref{background no matter}) and its solution, we obtain the following
perturbation equation in momentum space for RDE without matter from eq.(\ref{perturb einstein}) :
\begin{equation}\label{eom of phi2}
    \Phi_k''+ \left(3- \frac{1}{\alpha}\right)\Phi_k' -\frac{(2-\alpha)}{3\alpha}k^2H_0^{-2} e^{-2\left(\frac{1-\alpha}{\alpha}\right)x } \Phi_k   =   0
    \,.
\end{equation}
Here, the second term proportional to $\Phi_k'$ is a friction term, which will decrease the perturbation if $\alpha
>1/3$, while the third term will increase the perturbation. However, since $\alpha<1$, the exponential factor in this
term will make this term insignificant in the future ($x \gg 1$, or $a\gg1$). If $0<\alpha \le 1/3$, both terms will
increase the perturbation and make it unstable. Eq.(\ref{eom of phi2}) can be analytically solved by changing the
variable $\Phi_k = u_k e^{-\frac{1}{2}\left(3-\frac{1}{\alpha}\right)x}$ :
\begin{equation}\label{eom of u}
    \xi^2 \frac{d^2 u_k}{d\xi^2} + \xi \frac{d u_k}{d \xi} - \left[ k^2 H_0^{-2}\xi^2 + \left(\frac{3\alpha-1}{2-2\alpha}\right)^2 \right] u_k = 0
\end{equation}
where $\xi = \frac{\alpha}{1-\alpha}\left( \frac{2-\alpha}{3\alpha}\right)^{1/2} e^{-\frac{(1-\alpha)}{\alpha}x} \,$
and the solution of $\Phi_k$ is :
\begin{equation}
   \Phi_k = e^{-\frac{1}{2}\left(3-\frac{1}{\alpha}\right)x} \bigg(A_k I_\nu\left(k H_0^{-1}\xi\right) + B_k K_\nu\left(k H_0^{-1}\xi\right)\bigg) \,,
\end{equation}
where $\nu = |3\alpha-1|/(2-2\alpha)$, $ I_\nu$, $K_\nu$ are  modified Bessel functions and $A_k$, $B_k$ are
integration constants, which should be determined by the initial condition. When $x\gg1$, namely $a\gg1$, then $\xi \ll
1$ and if $\alpha > 1/3$
\begin{equation}
   \Phi_k(x\rightarrow \infty) \approx B_k e^{-\frac{1}{2}\left(3-\frac{1}{\alpha}\right)x} 2^{\nu -1}\Gamma(\nu)(k
   H_0^{-1}\xi)^{-\nu}  = B_k2^{\nu -1}\Gamma(\nu)(k H_0^{-1})^{-\nu}\left[\frac{1-\alpha}{\alpha}\left(
   \frac{3\alpha}{2-\alpha}\right)^{1/2}\right]^{\nu} \,,
\end{equation}
then for a given mode, the perturbation will be a constant in the far future. Otherwise, if $\alpha < 1/3$,
$\Phi_k(x\rightarrow \infty) \sim e^{-\left(3-\frac{1}{\alpha}\right)x}\rightarrow\infty$, and if $\alpha = 1/3$,
$\Phi_k(x\rightarrow \infty) \sim x \rightarrow\infty$. Therefore, only if $\alpha>1/3$, the perturbation is stable. We
also plot the evolution of perturbations $\Phi_k$ in Fig.\ref{fig::PhiK} and Fig.\ref{fig::PhiAlpha} for an
illustration. It should be noticed that $B_k$ should depending on $k$, so for a given initial condition of $\Phi_k$,
$B_k$ is different for each mode.

\emph{RDE with matter}: If the matter component also exist in the universe, the background equation (\ref{background no
matter}) becomes:
\begin{equation}\label{background with matter}
    h^2 = \frac{\rho_R + \rho_m}{3H_0^2} = \alpha \bigg[  \frac{(h^2)'}{2}  + 2h^2 \bigg] + \Omega_{m0}e^{-3x}\,,
\end{equation}
where $h \equiv H/H_0$, $\Omega_{m0} = \rho_{m0}/(3H_0^2)$ and we have used the energy conservation law for the matter.
The solution to the above equation is
\begin{equation}\label{background with matter sol}
    h^2 = \frac{2}{2-\alpha}\Omega_{m0}e^{-3x} +
    f_0e^{-\left(4-\frac{2}{\alpha}\right)x} \,,
\end{equation}
where $f_0$ can be determined by the condition $h(0)=1$:
\begin{equation}
    f_0 = 1-\frac{2}{2-\alpha}\Omega_{m0}\,.
\end{equation}
The energy density and pressure of RDE are
\begin{equation}
    \frac{\rho_R}{3H_0^2} = \frac{\alpha}{2-\alpha}\Omega_{m0}e^{-3x} +
    f_0 e^{-\left(4-\frac{2}{\alpha}\right)x} \,.
\end{equation}
and
\begin{equation}
    p_R = -\rho_R - \frac{\rho_R'}{3} = -\left(\frac{2}{\alpha}-1\right)f_0H_0^2e^{-\left(4-\frac{2}{\alpha}\right)x}
    \,.
\end{equation}
So, the equation of state at present is
\begin{equation}
    w_0 =
    -\frac{1}{3}\left(\frac{2}{\alpha}-1\right)\left(1-\frac{\alpha}{2-\alpha}\frac{\Omega_{m0}}{1-\Omega_{m0}}\right)
    \,.
\end{equation}
Again, only in the case of  $\alpha <1$, the universe will be accelerated expanding. Define
\begin{equation}
    \gamma \equiv \frac{\delta\rho_m}{\delta\rho_R} = \frac{c_{sR}^2}{c_s^2} - 1 = \left[\frac{\alpha}{2-\alpha} + \frac{2f_0}{3\Omega_{m0}}\left(2-\frac{1}{\alpha}\right)e^{\left(\frac{2}{\alpha}-1\right)x}\right]^{-1}\,,
\end{equation}
where $c_s^2 \equiv \delta p_R/\delta(\rho_R + \rho_m) = p_R'/(\rho_R' + \rho_m')$ is the total squared sound speed
while $c_{sR}^2 \equiv \delta p_R/\delta\rho_R = p_R'/\rho_R'$ is the squared sound speed for RDE only. By using the
background equation (\ref{background with matter}) and its solution (\ref{background with matter sol}), we obtain the
following perturbation equation for RDE with matter:
\begin{equation}\label{eom of phi2 case2}
   \Phi_k '' + \left[3-\frac{1}{\alpha}\left(\frac{2}{1+\gamma}-1+\beta\right)\right] \Phi_k'
   + \frac{2}{\alpha}\left(1-\frac{1}{1+\gamma}-\beta\right)\Phi_k
   -\frac{\lambda}{3\alpha}\left(\frac{2}{1+\gamma}-\alpha\right)k^2H_0^{-2} e^{-2\left(\frac{1}{\alpha}-1\right)x} \Phi_k   =   0 \,,
\end{equation}
where
\begin{equation}
    \beta \equiv \Omega_{m0}e^{-3x}h^{-2}  = \left[\frac{2}{2-\alpha} +
    \frac{f_0}{\Omega_{m0}}e^{\left(\frac{2}{\alpha}-1\right)x}\right]^{-1}\,,
\end{equation}
and
\begin{equation}
    \lambda \equiv \left[\frac{2}{2-\alpha} \Omega_{m0}e^{-\left(\frac{2}{\alpha}-1\right)x} + f_0\right]^{-1}\,.
\end{equation}
In the limit of $x \rightarrow \infty$ corresponding to the far future, then $\gamma,\beta \rightarrow 0$ and
$\lambda\sim \mathcal{O}(1)$, therefore, eq.(\ref{eom of phi2 case2}) comes back to eq.(\ref{eom of phi2}), which means
the matter perturbation is not so important for the perturbation of RDE in the far future. Since we are interesting in
the future behavior of the perturbations, we can make the following approximation:
\begin{equation}\label{approx}
    \gamma\approx \frac{3\Omega_{m0}}{2f_0}\left(\frac{\alpha}{2\alpha-1}\right)e^{-\left(\frac{2}{\alpha}-1\right)x}
    \,;\quad
    \beta\approx \frac{\Omega_{m0}}{f_0}e^{-\left(\frac{2}{\alpha}-1\right)x} \,;\quad
    \lambda \approx f_0^{-1} \;,
\end{equation}
and only keep linear terms of $\gamma$, $\beta$, then the perturbation equation (\ref{eom of phi2 case2}) becomes:
\begin{equation}\label{eom of u case2}
   u_k '' - \left[ \frac{(3\alpha-1)^2}{4\alpha^2}
   + \frac{(2\alpha^3+\alpha-3)\Omega_{m0}e^{-\left(\frac{2}{\alpha}-1\right)x}}{2\alpha^2(2\alpha-1)f_0}
   +\frac{2-\alpha}{3\alpha f_0}\left(1-\frac{3\alpha\Omega_{m0}e^{-\left(\frac{2}{\alpha}-1\right)x}}{(2-\alpha)(2\alpha-1)f_0}\right)k^2H_0^{-2} e^{-2\left(\frac{1}{\alpha}-1\right)x}\right]u_k   =   0 \,,
\end{equation}
where we have changed variable $\Phi_k$ to $u_k$ by
\begin{equation}\label{trans}
    \Phi_k = u_k \exp{\left[-\frac{1}{2}\left(3-\frac{1}{\alpha}\right)x - \frac{1}{2\alpha}\int\left(\frac{2\gamma}{1+\gamma}-\beta\right)dx\right]}
    \approx u_k \exp{\left[-\frac{1}{2}\left(3-\frac{1}{\alpha}\right)x \right]}\,.
\end{equation}
It seems that eq.(\ref{eom of u case2}) can not be solved analytically, but we can solve it in the long-wave and
short-wave limit as follows: For long-wave $(kH_0^{-1}\ll 1)$ modes, eq.(\ref{eom of u case2}) reduces to
\begin{equation}
    u_k '' + \left[
   \frac{(2\alpha^3+\alpha-3)\Omega_{m0}e^{-\left(\frac{2}{\alpha}-1\right)x}}{2\alpha^2(1-2\alpha)f_0} - \frac{(3\alpha-1)^2}{4\alpha^2}
   \right]u_k   =   0 \,,
\end{equation}
whose solution is
\begin{equation}
    u_k= C_{1k}J_\nu(\xi_1) + C_{2k}Y_\nu(\xi_1)\,,
\end{equation}
where $\nu = |3\alpha-1|/(2-\alpha)$, $ J_\nu$, $Y_\nu$ are Bessel functions, $C_{1k}$, $C_{2k}$ are integration
constants and
\begin{equation}
    \xi_1 = \frac{2\alpha}{2-\alpha}\left[\frac{(2\alpha^3+\alpha-3)\Omega_{m0}}{2\alpha^2(1-2\alpha)f_0}\right]^{1/2}
    e^{-\frac{1}{2}\left(\frac{2}{\alpha}-1\right)x} \,.
\end{equation}
Therefore, when $x\rightarrow\infty$ and $\alpha >1/3$, the perturbation $\Phi_k(x\rightarrow\infty) \approx
\Phi_c(k)$, where $\Phi_c(k)$ is a constant for a given mode. Otherwise, if $\alpha = 1/3$, $\Phi_k(x\rightarrow
\infty) \sim e^{-\left(3-\frac{1}{\alpha}\right)x}\rightarrow\infty$,  and $\Phi_k(x\rightarrow \infty) \sim x
\rightarrow\infty$ if $\alpha = 1/3$. Therefore, only if $\alpha>1/3$, the perturbation is stable in the long-wave
limit.

\begin{figure}[h]
\includegraphics[width=0.4\textwidth]{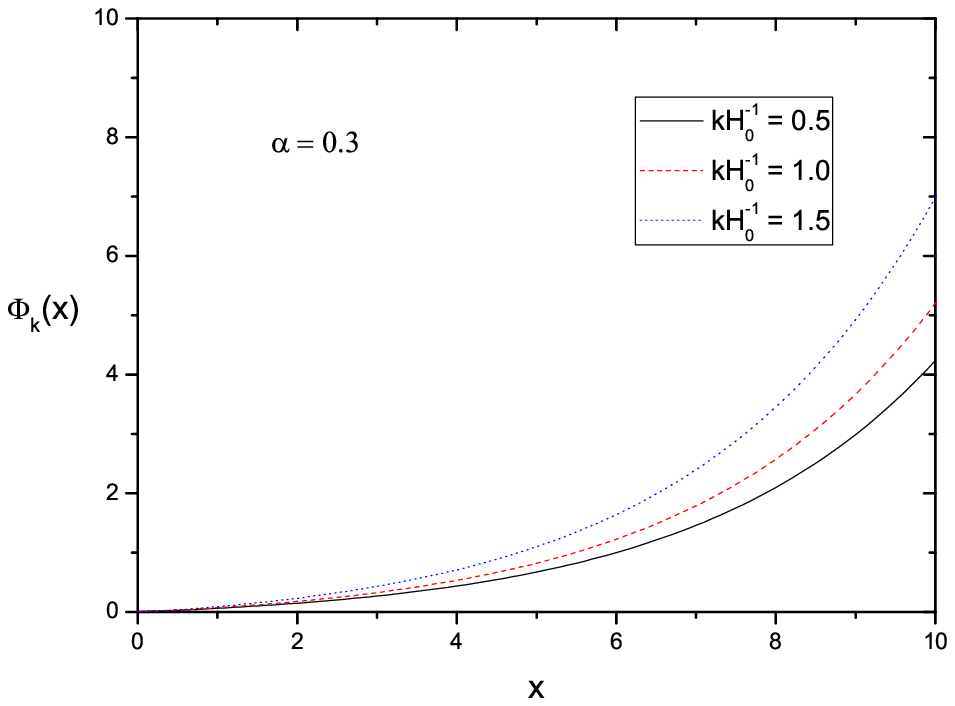}
\quad
\includegraphics[width=0.4\textwidth]{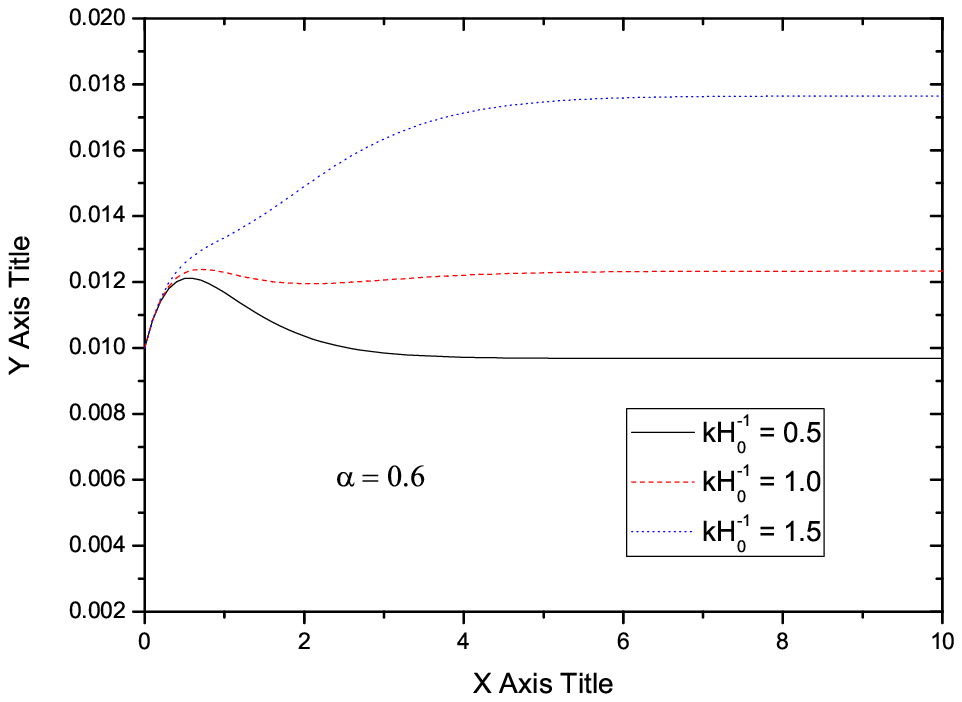}
\caption{\label{fig::PhiK2}\emph{RDE with matter}: Evolution of perturbations $\Phi_k$  with $\alpha = 0.3$(left),
$0.6$(right), $kH_0^{-1} = 0.5, 1.0, 1.5$ and the initial condition $\Phi_{k0} = 0.01$ and $\Phi'_{k0} = 0.01$.}
\end{figure}

\begin{figure}[h]
\includegraphics[width=0.31\textwidth]{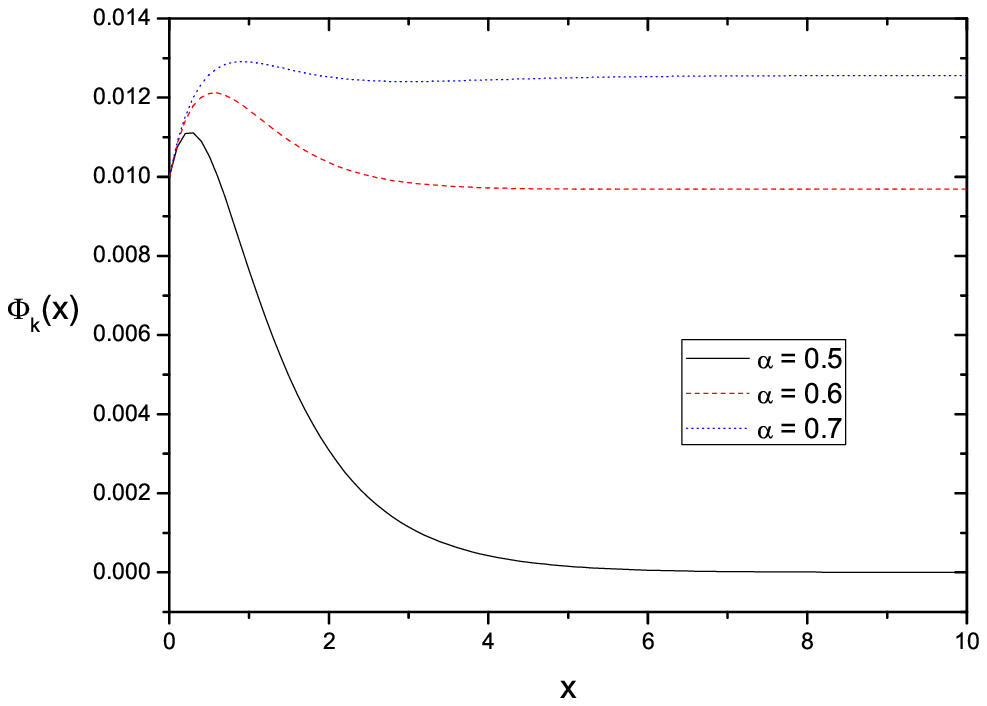}
\quad
\includegraphics[width=0.31\textwidth]{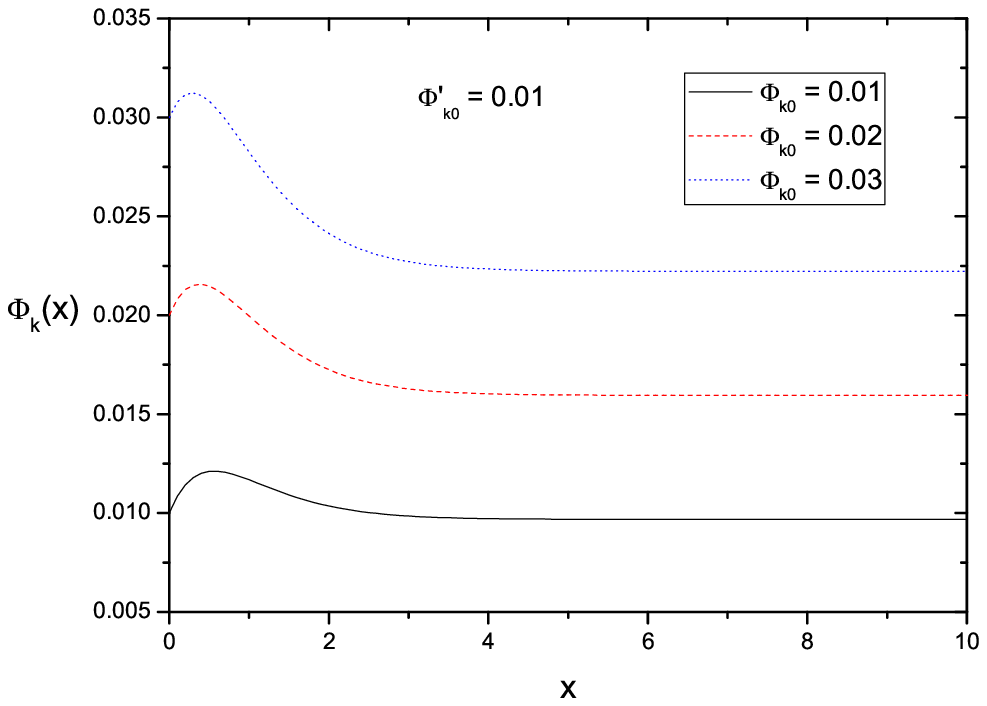}
\quad
\includegraphics[width=0.31\textwidth]{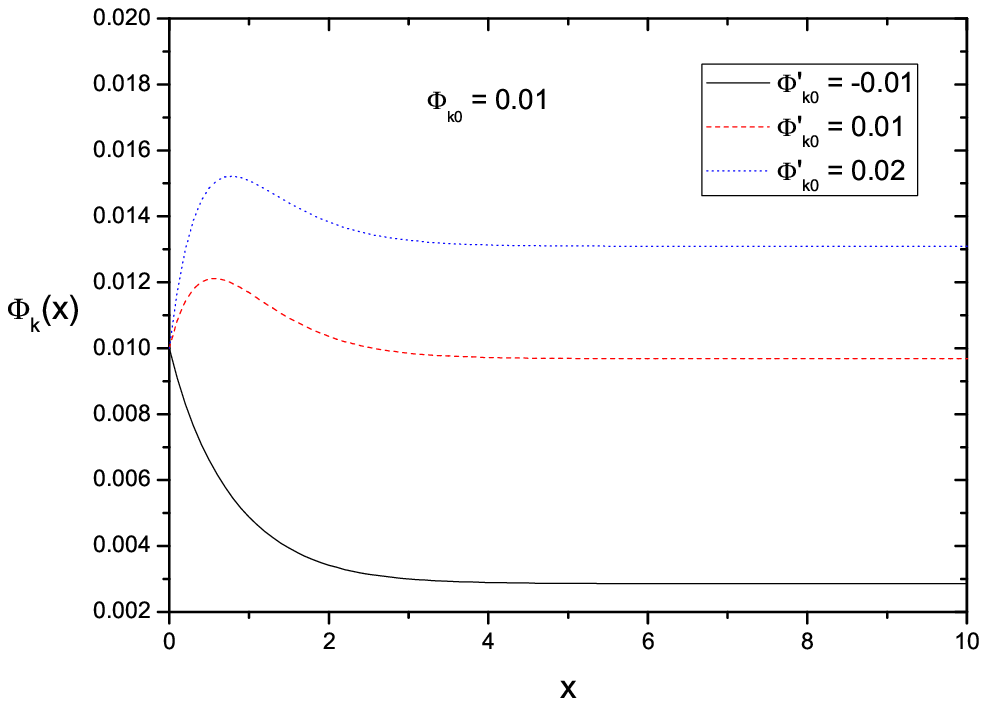}
\caption{\label{fig::PhiAlpha2}\emph{RDE with matter}: The most left figure shows the evolution of perturbations
$\Phi_k$ with $\alpha = 0.5, 0.6, 0.7$ and the initial condition $\Phi_{k0} = 0.01$, $\Phi'_{k0} = 0.01$. The middle
and the most right figures indicate the evolution of $\Phi_k$ with $\alpha = 0.6$ and different initial conditions:
$\Phi_{k0} = 0.01, 0.02, 0.03$, $\Phi'_{k0} = 0.01$ in the middle one and $\Phi_{k0} = 0.01$, $\Phi'_{k0} = -0.01,
0.01, 0.02$ in the most right one.}
\end{figure}

For short-wave $(kH_0^{-1}\gg 1)$ modes, eq.(\ref{eom of u case2}) reduces to
\begin{equation}
    u_k''-\left[ \frac{2-\alpha}{3\alpha f_0}k^2H_0^{-2} e^{-2\left(\frac{1}{\alpha}-1\right)x}\right]u_k   =   0 \,,
\end{equation}
and here we have neglected the second term proportional to $e^{-\left(\frac{2}{\alpha}-1\right)x}$, since it is small
in the far future($x \rightarrow \infty$) and its solution is:
\begin{equation}
    u_k = D_{1k} I_0\left(k H_0^{-1}\xi_2\right) + D_{2k} K_0\left(k H_0^{-1}\xi_2\right)\,.
\end{equation}
so, $u_k(x\rightarrow\infty) \sim x $. Here  $D_{1k}$, $D_{2k}$ are integration constants and
\begin{equation}
    \xi_2 = \frac{\alpha}{1-\alpha}\left(\frac{2-\alpha}{3\alpha f_0}
    \right)^{1/2}e^{-\left(\frac{1}{\alpha}-1\right)x} \,.
\end{equation}
Again, the perturbation is stable if $\alpha > 1/3$ in the short-wave limit. If $\alpha = 1/3$,
$\Phi_k(x\rightarrow\infty)\sim x \rightarrow \infty$ and if $\alpha < 1/3$, $\Phi_k(x\rightarrow\infty)$ is
exponential increased. We also numerically solve the eq.(\ref{eom of phi2 case2}) and plot the evolution of $\Phi_k$ in
Fig.\ref{fig::PhiK2} and Fig.\ref{fig::PhiAlpha2}.

In conclusion, we have study the scalar perturbation of the Ricci dark energy (RDE) with and without matter in the flat
FRW universe. To make the universe accelerated expanding, it requires the parameter $\alpha < 1$ in RDE, and only if
$\alpha > 1/3$, the scalar perturbation is stable. Therefore, from theoretical aspect, the reasonable value for
$\alpha$ is $ 1/3< \alpha < 1$. According to the joint analysis in ref.\cite{Zhang:constraint}, the best-fit results
with $1\sigma$ uncertainly is $\alpha = 0.359_{-0.025}^{+0.024}$ and our result indeed gives a lower bound of $\alpha$
theoretically, which is more stringent than that from observations.

\begin{acknowledgments}
This work is supported by National Science Foundation of China grant No. 10847153 and No. 10671128.
\end{acknowledgments}

\end{document}